\documentclass{moriond}

\def\Journal#1#2#3#4{{#1} {\bf #2}, #3 (#4)}

\def\be{\begin{equation}}
\def\ee{\end{equation}}
\def\bea{\begin{eqnarray}}
\def\eea{\end{eqnarray}}

\newcommand{\Photo}{}

\begin{document}
\vspace*{4cm}
\title{Reconstruction of very inclined air showers with radio data 
and application to the first GRANDProto300 cosmic-ray candidates}

\author{Marion Guelfand*, Valentin Decoene, Olivier Martineau, Simon Prunet, Sebastian Castro-Isern, Nathan Lebas, on behalf of the GRAND collaboration}

\address{*CPPM, 163, avenue de Luminy
Case 902
13288 Marseille cedex 09
}

\maketitle\abstracts{Radio detection of very inclined extensive air showers (EAS) offers a promising way to study ultra-high-energy cosmic rays (UHECRs) and neutrinos above $10^{17}$ eV. We present a method to reconstruct the arrival direction and energy of EAS with zenith angles above $60^\circ$, applicable in principle to upward-going neutrino-induced showers. The approach combines a point-source-like description of the radio wavefront with the Angular Distribution Function (ADF), a phenomenological model describing the angular pattern of electric field amplitudes in the 50-200 MHz band. On realistic cosmic-ray simulations, we achieve better than $0.07^\circ$ resolution on arrival direction and ~15\% on primary energy. Applied directly to voltage traces, this method allows efficient cosmic-ray selection with accurate direction reconstruction and first-order energy estimation. We illustrate the method on the first cosmic-ray candidates detected by the GRANDProto300 prototype of GRAND, validating its detection principle and paving the way for high-precision neutrino searches.}

\section{Introduction}
Ultra-high-energy (UHE) cosmic rays and neutrinos can be detected via the radio emission produced by extensive air showers. Radio arrays are robust, cost-effective, and scalable, making them well suited for covering large areas and probing low fluxes. Future radio experiments focus on very inclined showers, as (i) UHE neutrinos detected in air have Earth-skimming trajectories and (ii) such inclined showers produce large radio footprints, allowing sparse detector layouts.

In this context, GRANDProto300 (GP300)~\cite{Martineau_2025}, a prototype of the GRAND project~\cite{Guépin_2026}, is currently being commissioned. Its goal is to validate autonomous radio detection by measuring cosmic rays in the $10^{16.5}$–$10^{18}$ eV range and reconstructing their properties. New reconstruction methods, such as the Angular Distribution Function (ADF), are directly tested on experimental data. In this work, we briefly present the ADF method and demonstrate its performance for event selection and reconstruction (direction and energy).

\section{The ADF model}
The ADF method aims at achieving sub-degree angular resolution for very inclined showers, a key requirement for neutrino astronomy and multi-messenger studies. 
It relies on two observables: the signal arrival time and the peak electric field amplitude. The method was initially developed at the electric-field level~\cite{Decoene_2021}~\cite{Guelfand_2025}, rather than using voltage signals, which avoid dependence on detector response, and focuses on the intrinsic physics of the radio emission. It is tailored for the 50–200 MHz frequency band used by GRAND.
Two pieces of information are needed to determine the full shower geometry: the shower axis and a reference emission point $X_\textrm{source}$. The latter assumes that the radio emission can be treated as point-like, an approximation that is valid for inclined showers~\cite{Decoene_2022}.
The reconstruction proceeds in two steps. First, $X_\textrm{source}$ is determined by a spherical fit of the radio wavefront using trigger times, taking atmospheric refraction into account. Second, the shower axis is reconstructed by fitting the ADF function to the measured peak amplitudes $A_i$:
$R(\theta, \phi) = \sum_{i=1}^{N_{\rm ant}} \left[ A_i - f_i^{\rm ADF}(A, \theta, \phi, \delta \omega; X_e) \right]^2$,
\vspace{-0.1cm}
with ADF model $f_i^{\rm ADF}$:
\vspace{-0.1cm}
\begin{equation}
\label{eq:adf}
f^{\rm ADF} (A, \theta, \phi, \delta \omega; X_e) = \frac{A}{l} \left[ 1 + G_A \frac{\cos \eta}{\sin \alpha} \right] \cdot \frac{1}{1 + 4 \left( \frac{\tan \omega / \tan \omega_c - 1}{\delta \omega} \right)^2 } \ ,
\end{equation}
\vspace{-0.1cm}
Here, $\omega$ and $\eta$ are the radial and angular antenna coordinates in the shower plane, $\omega_c$ the Cherenkov angle, which is computed independently 
using a simple toy model, $l$ the distance to $X_e$, and $G_A$ the geomagnetic asymmetry factor. 
The fit determines the shower direction $(\theta, \phi)$, a scaling factor $A$, and the angular width $\delta \omega$.

Finally, the electromagnetic energy is estimated from $A$, corrected for geomagnetic and atmospheric effects~\cite{Guelfand_2025}: $E^{*}_{\rm em} = \frac{A}{\sin(\alpha) \, f(\rho, \sin(\alpha))}$ with $f(\rho, \sin\alpha)$ derived from simulations.

Performances on realistic GP300 simulations~\cite{Guelfand_2025} show angular resolution at $0.07^\circ$ and energy resolution at less than $15\%$ on the primary.

\section{Reconstruction of GRAND cosmic-ray candidates}
\subsection{Commissioning data from GP300}
The GP300 array is currently in its commissioning phase, with 65 out of the planned 289 antennas deployed as of January 2025. Efforts focus on detector calibration, thermal regulation, and suppression of radio self-emission, while the antenna-level trigger is being optimized. Since December 2024, events are recorded when at least five antennas trigger within a 10 $\mu s$ coincidence window. After initial filtering~\cite{Lavoisier_2026}, about 18 000 preliminary cosmic-ray candidates were identified between July and October 2025. The ADF method is then applied, followed by post-reconstruction selection based mainly on amplitude information. This criterion is crucial, as EAS exhibit a characteristic profile with a strong enhancement at the Cherenkov ring, which is not reproduced by background signals.

\subsection{Selection on raw ADC data}
In the following, the ADF reconstruction is applied directly on raw detector voltage signals, bypassing the electric-field reconstruction step and allowing direct event identification from raw data. Although ADF was initially designed for electric-field data, this approach has been validated on cosmic-ray simulations~\cite{Guelfand_2025_ICRC}, demonstrating sub-degree angular resolution and providing a first proxy for energy.

After reconstruction, a series of selection cuts are applied:
\begin{enumerate}
    \item Zenith angle: Only events with $\theta < 80^\circ$ are kept, avoiding large background near the horizon and complex ground effects. This cut will be removed in the future.
    \item Angular distance to shower axis: Events are rejected if any triggered antenna has an angular distance $\omega \geq 3^\circ$ from $X_\textrm{source}$. This ensures that the event is consistent with the expected beaming: all triggered antennas must lie within or close to the Cherenkov cone ($\omega_c\sim 0.4^\circ-1^\circ$) depending on zenith angle~\cite{Guelfand_2025}, due to relativistic effects. Any antenna beyond $3^\circ$ likely measures background or a misreconstructed signal, and thus the event is discarded.
    \item Fit quality: Events with reduced $\chi^2_{\nu, \textrm{ADF}} < 20$ are retained. In simulations, $\chi^2_{\nu, \textrm{ADF}}$ typically averages $\sim 5$~\cite{Guelfand_2025_ICRC}, so the cut safely includes realistic variations without rejecting valid events. The relatively high value of the reduced $\chi^2_{\nu, \textrm{ADF}}$ value is due to the fact that the model is applied to voltage rather than electric-field signals, and that the amplitude uncertainty $\sigma_A$ is conservatively set to 7.5\% as done for AERA~\cite{Auger_2017}, although it has not yet been measured for our detector.
\end{enumerate}

After these cuts, 51 high-quality cosmic-ray events are selected. This number is lower than the expected rate ($\sim$ 20/day with this setup over $\sim$ 1500 hours of data taking) because the focus here is on validating hardware and reconstruction, not yet on full physics studies. These candidates are in addition to the 26 events reported at ICRC2025~\cite{Lavoisier_2025}.

\subsection{Arrival direction and energy of the selected candidates}
\begin{figure}[!h]
\begin{center}
        \includegraphics[width=0.5\columnwidth]{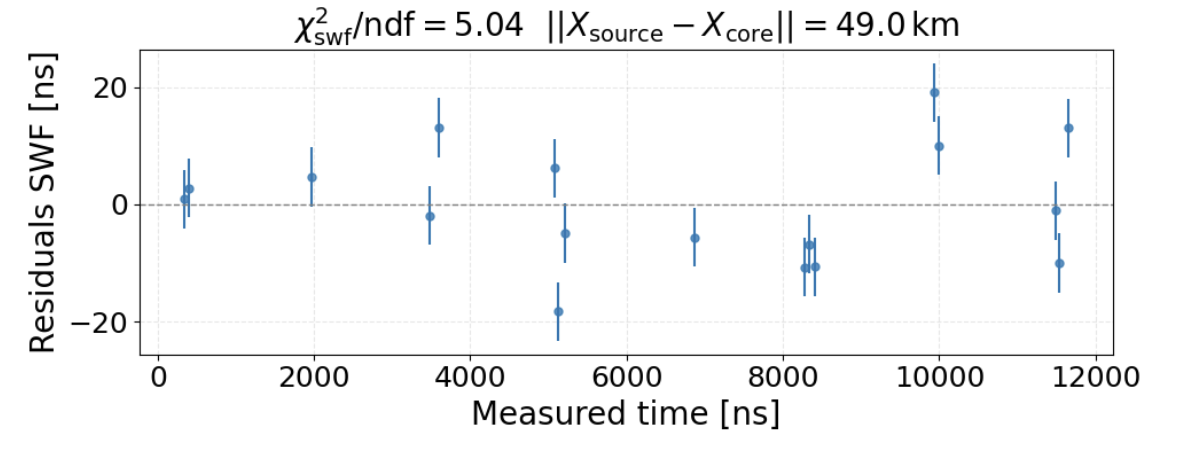}
    \end{center}
         \includegraphics[width=0.5\columnwidth]{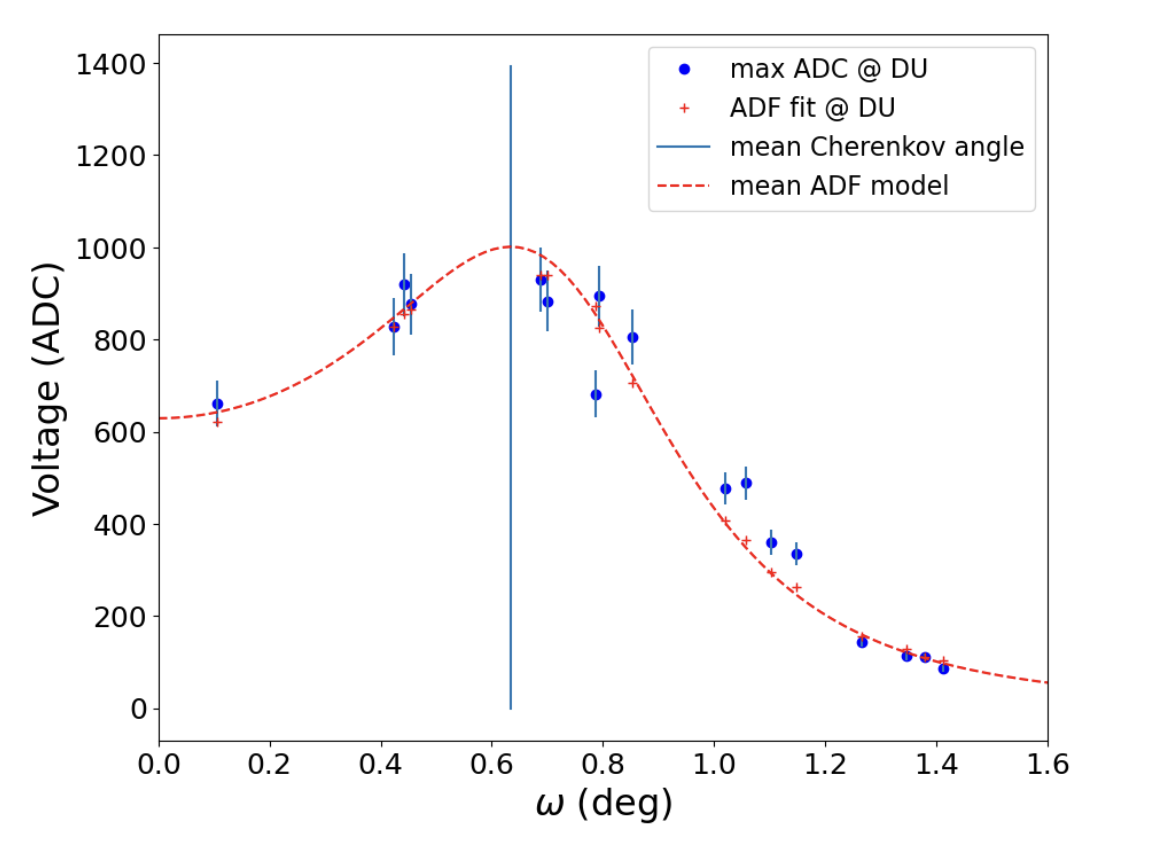}
         \includegraphics[width=0.48\columnwidth]{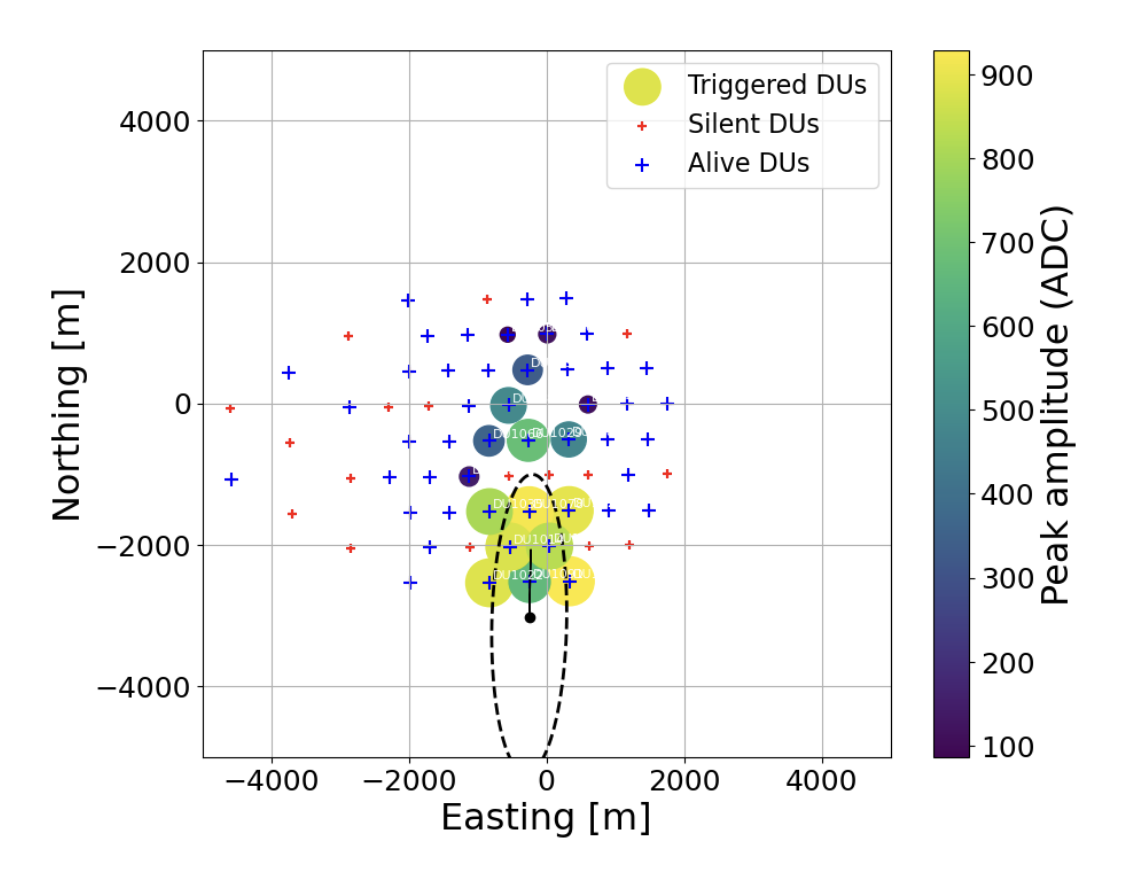}
    \caption{All three panels correspond to the same event, selected among the 51 candidates identified in the data set from July to October 2025. \textit{(Top)} Time residuals of the spherical fit for triggered antennas. The error bars correspond to an estimated timing uncertainty of $\sigma_t = 5$ ns. 
\textit{(Bottom left)} Amplitude profile: measured peak amplitudes (blue) and ADF model (red), with $\sigma_A = 7.5\%$ (error bar). The mean fitted profile (red line) and reconstructed $\omega_c$ (blue line) are shown. 
\textit{(Bottom right)} GP300 array layout: active (blue) and inactive (red) detection units (DUs), with triggered antennas (color circles), reconstructed core, and footprint.}
\label{fig:candidate1}
\end{figure}

One cosmic-ray candidate is presented to illustrate the validation of the method. Figure~\ref{fig:candidate1} (top) shows the time residuals of the triggered antennas as a function of the measured arrival time, with a reduced $\chi^2_{\nu,\textrm{SWF}} \sim 5$. This relatively large value is mainly due to the fact that the time calibration is not yet finalized and a nominal timing uncertainty of 5 ns is currently assumed. However, the associated uncertainty primarily affects the longitudinal position of the emission point, while the lateral resolution remains high and thus does not impact the direction reconstruction. The distance between the reconstructed emission point $X_\textrm{source}$ and the shower core $X_\textrm{core}$ is about 49 km, consistent with simulations. At this stage, the nature of the primary particle is not yet reconstructed, although ongoing work aims to exploit the $X_\textrm{source}$ information for this purpose.

Figure~\ref{fig:candidate1} (bottom left) shows the amplitude profile as a function of the angular distance $\omega$. The ADF fit (red line) yields $\chi^2_{\nu,\textrm{ADF}} = 4.7$, a direction $\theta = 75.1^\circ \pm 0.03^\circ$, $\phi = 358.7^\circ \pm 0.02^\circ$, and an energy $E = 2.08 \times 10^{18}$ eV. The profile exhibits the expected enhancement around the Cherenkov angle ($\omega_c \sim 0.6^\circ$) and a rapid decrease outside the ring.

Figure~\ref{fig:candidate1} (bottom right) shows the layout of the 65 antennas of the GP300 array, where colored markers indicate triggered antennas. Some antennas (shown as red crosses) were not operational during data taking. The reconstructed shower core position and footprint are also displayed. The elongated footprint is a clear signature of a cosmic-ray event, in contrast with the more isotropic patterns expected from background.

\begin{figure}[!h]
         \includegraphics[width=0.45\columnwidth]{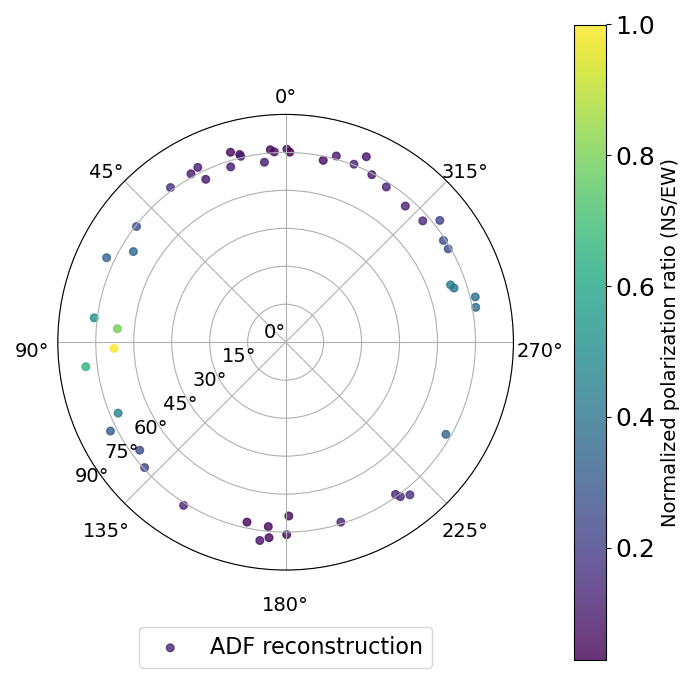}
         \includegraphics[width=0.5\columnwidth]{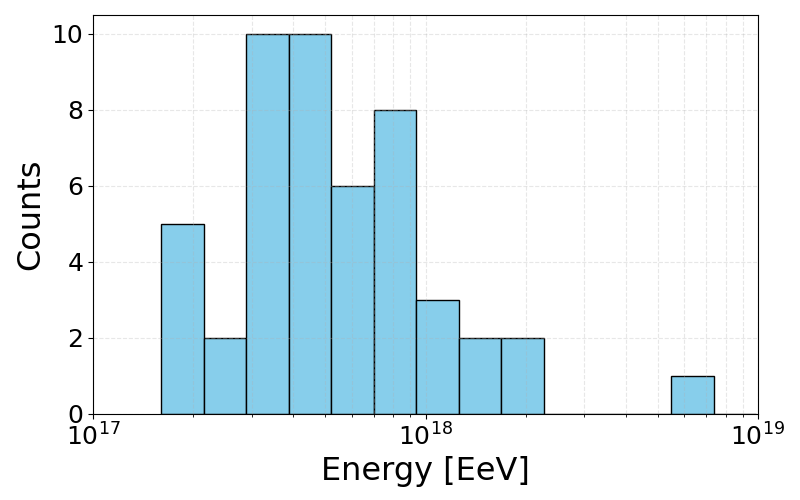}
    \caption{\textit{(Left)} Reconstructed arrival directions of the 51 cosmic-ray candidates from ADC data. The color scale shows the normalized ratio of mean peak amplitudes in the North–South and East–West arms. North–South signals correspond to East–West arrivals, and vice versa. 
\textit{(Right)} Reconstructed energy distribution of the 51 candidates. Calibration and reconstruction procedures are still being refined.}
      \label{fig:allcandidates}
\end{figure}

Figure~\ref{fig:allcandidates} (left) shows the sky distribution of arrival directions. The observed North excess compared to South is consistent with expectations from the geomagnetic field at the GP300 site. Figure~\ref{fig:allcandidates} (right) presents the reconstructed energy distribution, already in agreement with expectations despite being derived from raw ADC data. 
Independent reconstruction methods provide cross-checks of the results~\cite{Ferrière_2026} \cite{Gulzow_2024}, ensuring robustness. In particular, the reconstructed energy distribution is consistent with that reported in~\cite{Ferrière_2026}.

Overall, these results demonstrate that autonomous radio detection can reliably identify cosmic-ray events and reconstruct both their direction and energy. The ADF model proves to be a powerful tool for these analyses, and the commissioning phase is expected to conclude by the end of 2026.

\section*{References}

\end{document}